\documentclass[prc,unsortedaddress,superscriptaddress,twocolumn,amsmath,amsfonts,amssymb,showpacs,floatfix]{revtex4}

\begin{document}
\title{Unitary correlation in nuclear reaction theory}

\author{A. M. Mukhamedzhanov} 
\affiliation{Cyclotron Institute, Texas A\&M University, College
Station, TX 77843}

\author{A. S. Kadyrov}
\affiliation{ARC Centre for Antimatter-Matter Studies, Curtin University,
GPO Box U1987, Perth, WA 6845, Australia}

\begin{abstract}
We prove that the amplitudes for the $(d,p),\,(d,pn)$ and $(e,e'p)$ reactions determining the asymptotic behavior of the exact scattering wave functions in the corresponding channels are invariant under unitary correlation operators while the spectroscopic factors are not. Moreover, the exact reaction amplitudes are not parametrized in terms of the spectroscopic factors and cannot provide a tool to determine the spectroscopic factors. 
\end{abstract}
\pacs{24.10.Cn, 24.10.Eq, 25.70.Ef, 13.75.Cs}
\today
\maketitle

{\em{Introduction.}}---Since the dawn of nuclear physics $(d,p)$ reactions have been the main tool for extracting spectroscopic factors (SFs) (we call them phenomenological SFs), which were compared with predictions of the  independent-particle shell model (ISPM) (see \cite{lee06} and references therein). Later on electron-induced breakup reactions (see \cite{kramer} and references therein) and nucleon knockout reactions (see \cite{gade} and references therein) became new tools to determine the SFs. Reduction of the phenomenological SFs  deduced from the $(e,e'\,p)$ reactions compared to the IPSM ones has been discussed in \cite{kramer}. Recently a similar reduction has been observed for the SFs determined from the analysis of the single-nucleon knockout nuclear reactions \cite{gade} as well. 
There are different possible sources of this reduction: the single-particle approximation to the overlap function, antisymmetrization effects, an ambiguity of the optical potentials, accuracy of the DWBA, contribution of the coupled channels or the effect of the short-range $NN$ correlations.
 
In the IPSM  the model wave functions are given by the linear combination of the Slater determinant wave functions. However, the such model wave functions don't take into account short-range $NN$ correlations  caused by the repulsive core in the $NN$ potential. In \cite{feldmeier} the so-called unitary correlation operator method (UCOM) has been developed, which allows one to correct the IPSM taking into account the short-range repulsive $NN$ correlations by applying unitary correlation operators (UCOs) onto the trial functions. In this work we consider the effect of such short-range correlations and demonstrate that the exact reaction amplitude is invariant under finite-range unitary transformation of the wave functions. 

{\em{$(d,\,p)$ reaction amplitude.}}---We start our consideration from the deuteron stripping reaction $A(d,p)B$. Its amplitude in the post-form is given by
\begin{equation}
M_{dp} =  <\chi_f^{( - )} \varphi_{B}|V_{pB}  - U_{pB}|\Psi_{i}^{( + )}>.
\label{ampldp1}
\end{equation}
Here $\Psi_{i}^{( + )}$ is the exact $d+A$ scattering wave function which has outgoing waves in all the open channels and satisfies
\begin{align}
(T_{pB}+T_{B}+V_{pB}+V_{B}) \Psi_{i}^{( + )} = E \Psi_{i}^{( + )} ,
\label{Psi1}
\end{align}
$\chi_{f}^{(-)}$ is the distorted wave in the exit $p+ B$ channel. 
It is a solution to
$(T_{pB}+T_{B}+U_{pB}+V_{B}) \chi_{f}^{( - )} = E \chi_{f}^{(-)}$, 
where $T_{pB}$ is the kinetic energy operator for the relative motion of the proton and the center of mass of nucleus $B$ while $T_{B}$ is the kinetic energy operator for internal motion of the nucleons 
in nucleus $B$. In addition, $\varphi_{B}$ is the nucleus $B$ bound state wave function and $V_{ij}$ is the interaction potential of nuclei $i$ and $j$, $U_{ij}$ is their optical potential, $V_{B}$ is the interaction potential of the nucleons in nucleus $B$. Note that the potential $V_{pB}$ may consist of the sum of the two- and three-nucleon forces. 
In the paper we will take into account only two-nucleon forces, i.e. $V_{pB}= V^{N}_{pB} + V^{C}_{pB}$,  
$\,V_{pB}^N  = \sum\limits_i{V_{pi}^N }$, where $V_{pi}^{N}$ is the proton-nucleon interaction potential with a repulsive core. 
We note that the initial scattering wave function $\Psi_{i}^{( + )}$ is fully antisymmetrized, while in the final channel only the bound state wave function $\varphi_{B}$ is antisymmetrized. 
The transition operator is symmetric relative to the nucleon interchange what can be easily seen by rewritting it in terms of the kinetic energy operators (see Eq. (\ref{ampldpsurf1}) below). 

Eq. (\ref{ampldp1}) represents a conventional expression for the $(d,p)$ reaction in terms of the volume integral. Recently we have developed a new surface-integral based formulation of the reaction theory with charged particles (see \cite{kadyrov09} and references therein). Transition to the surface integral in Eq. (\ref{ampldp1}) can be achieved by replacing the potentials in the transition operators by the kinetic energy operators. Below we demonstrate it in a few simple and transparent steps. 
First we rewrite  Eq. (\ref{ampldp1}) as   
\begin{align}
M_{dp} = & <\chi_f^{( - )} \varphi_{B}|V_{pB} + V_{B} - U_{pB} - V_{B}|\Psi_{i}^{( + )}>  \nonumber\\
= & <\chi_f^{( - )} \varphi_{B}|\overleftarrow T - \overrightarrow T| \Psi_{i}^{( + )}>, 
\label{ampldpsurf1}
\end{align}  
where $\overrightarrow T$  ($\overleftarrow T$) the total kinetic energy operator acting to the right (left) and  $T=\sum\limits_{i}T_{i}$.

Using Eq. (\ref{ampldpsurf1}) we demonstrate that the exact reaction amplitude $M_{dp}$ is invariant 
under finite-range unitary transformation of the wave functions. Such transformation has been developed in \cite{feldmeier} to take into account short-range nucleon-nucleon correlations. 
The problem is that in the IPSM, which is used to calculate SFs,
the nucleus wave function is given by the linear combination of the Slater determinants. However, 
finite number of the Slater determinant components cannot reproduce depletion of the nucleus wave function generated by a strong short-range repulsive core in the $NN$ interaction, i.e. the exact wave function contains components in the subspace $Q$ beyond of the model space covered by the IPSM components.  As mentioned above, the unitary correlation operator method was developed \cite{feldmeier} in order to take into account effectively  the repulsive core in the $NN$-interaction.
The correlated wave function in the UCOM can be obtained from an uncorrleated state (linear combination of Slater determinants)  using the unitary correlator $\Psi= {\hat C} {\tilde \Psi}$.
Here, ${\tilde \Psi}$ is an uncorrelated state, ${\hat C}$ is the unitary correlation operator (UCO) given by \cite{feldmeier} ${\hat C}= e^{-i\,G}$, where $G$ consists of the radial and tensor terms and can be written as $G=\sum\limits_{i < j}\,g(ij)$. In this paper we restrict ourselves only by the two-body operators $g(ij)$ assuming that the higher order terms give negligible contribution, although the final result is valid for higher order correlations as well. Note that the two-body correlator can generate three- and higher order potential terms even if the initial potentials are two-body ones \cite{feldmeier}. By definition each $g(i\,j)$ depends on the distance between nucleons $i$ and $j$. It is designed to shift away nucleons from the classically forbidden region ($\approx 1-1.5$ fm) generated by the repulsive core and disappears in the clasically allowed region, i.e. $g(i\,j)$ is a short-range correlator quickly going to zero as distance between the correlated nucleons increases. Similarly three-body correlator disappears when one of the three correlated nucleons is moved away from the correlated volume.  
 
Evidently, the application of the UCO conserves the norm  $<\Psi|\Psi>\,= \,<{\tilde {\Psi}}|{\tilde {\Psi}}>$ and the asymptotic normalization coefficient (ANC), which determines the amplitude of the tail of the radial overlap function of nuclei $A$ and $B=(A\,n)$ (for simplicity we consider the neutron ANC):
\begin{eqnarray}
I_{A}^{B}(r)= <\varphi_{A}|\varphi_{B}>      
 \stackrel{r \to \infty}{
\approx}  C_{An}^{B}\,\frac{e^{-\kappa\,r_{nA}}}{r_{nA}},
\label{overlfunct1}
\end{eqnarray}
where integration in the matrix element $<\varphi_{A}|\varphi_{B}>$ is carried over all the internal coordinates of daughter nucleus $A$; $\kappa$ is the neutron bound state wave number and ${\bf {\rm r}}_{nA}$ is the radius connecting $n$ and the center-of-mass of nucleus $A$. Assuming now that both wave functions $\varphi_{B}$
and $\varphi_{A}$ are correlated we get from Eq. (\ref{overlfunct1}) that
\begin{align}
I_{A}^{B}(r_{nA})=& <\varphi_{A}|\varphi_{B}> =  <{\tilde \varphi_{A}}| {\hat C}_{A}^{-1}\,{\hat C}_{B}|{\tilde \varphi_{B}}>                            \nonumber\\
= &<{\tilde \varphi_{A}}|{\hat C}_{nA}|{\tilde \varphi_{B}}>
  \stackrel{r_{nA} \to \infty}{\approx}  C_{An}^{B}\,\frac{e^{-\kappa\,r_{nA}}}{r_{nA}},
\label{overlfunct2}
\end{align}
where we used the cluster property of the UCO  
\begin{equation}
 {\hat C}_{B}= {\hat C}_{A}\,{\hat C}_{nA}.
\label{claster1}
\end{equation}  
Here, $C_{B}$ is the UCO between nucleons of nucleus $B$ and $C_{nA}$ is the UCO between neutron and nucleons of nucleus $A$.

From the definition of the SF 
\begin{equation}
S= <I_{A}^{B}|I_{A}^{B}> 
\label{sf1}
\end{equation}
we can conclude that the SF, in contrast to the ANC, is not invariant under UCO. 
 
Now using the surface-integral formulation for the reaction amplitude we will prove that 
the reaction amplitude is invariant under short-range unitary transformation of the wave functions. Let $\Psi_{i}^{(+)}$ to be the correlated wave function in the initial channel (i.e. it contains the incident $d+A$ wave), which is related to the uncorrelated state $\Psi_{i}^{(+)}={\hat C}\,{\tilde \Psi}_{i}^{(+)}$.
Correspondingly the correlated channel wave function in the final state  $\Phi_{f}^{(-)}= \varphi_{B}\,\chi_{f}^{(-)}$  can be written in terms of the uncorrelated one as 
\begin{equation}
\Phi_{f}^{(-)}= {\hat C}_{B}\,{\tilde \Phi}_{f}^{(-)} = {\hat C}_{B}\,{\tilde \varphi}_{B}\,\chi_{f}^{(-)}.
\label{Phif1}
\end{equation} 
Now we can rewrite   
\begin{eqnarray}
M_{dp} = <\chi_f^{( - )}{\tilde  \varphi}_{B}\,{\hat C}_{B}^{-1}|(\overleftarrow T - \overrightarrow T)|{\hat C}\,{\tilde \Psi}_{i}^{( + )}>, 
\label{amplcorr1}
\end{eqnarray}
Note that the distorted wave $\chi_f^{( - )}$ cannot be considered as a correlated wave function because it is a solution of the Schr\"odinger equation with the optical potential, which depends on the distance between the proton and the center-of-mass of nucleus $B$ and doesn't have short-range repulsive core. 
We can distinguish between the outgoing proton and nucleons belonging to $B$  because there is no antisymmetrization between the outgoing proton and nucleons of nucleus $B$, i.e. one can tag the outgoing proton. 
Once the transition operator is expressed in terms of the difference of the kinetic energy operators we can apply the Green's theorem and transform the volume integral into the surface one \cite{kadyrov09}. We rewrite the total kinetic energy operator $T$ as $T= T_{pB} + T_{B}$.
Since in the bra state we have the bound state wave function of nucleus $B$ the transformation of the volume integral over internal coordinates of nucleus B and extending them to infinity
leads to the disappearance of the term $<  |(\overleftarrow T_{B} - \overrightarrow T_{B})| > $. Another equivalent way to prove that  $<  |(\overleftarrow T_{B} - \overrightarrow T_{B})| >$ disappears is to use the fact that,
due of the presence of the bound state wave function of nucleus $B$ in the bra state, the operator $T_{B}$ is hermitian. Therefore, taking twice the integration by parts we can transform $\overleftarrow T_{B}$ to $ \overrightarrow T_{B}$  resulting into expression
\begin{eqnarray}
M_{dp} = < {\tilde \Phi}_{f}^{(-)}|(\overleftarrow T_{pB} - \overrightarrow T_{pB})\,{\hat C}_{pB}|{\tilde \Psi}_{i}^{( + )}>.  
\label{amplcorr2}
\end{eqnarray} 
Now we split the volume integral (\ref{amplcorr2}) into two parts:
\begin{align}
M_{dp} = &< {\tilde \Phi}_{f}^{(-)}|(\overleftarrow T_{pB} - \overrightarrow T_{pB}){\hat C}_{pB}|{\tilde  \Psi}_{i}^{( + )}> \nonumber\\ 
=  & {M_{dp}}|_{r_{pB} \le R} + {M_{dp}}|_{r_{pB} > R}. 
\label{reactampl3}
\end{align}
Here ${\rm {\bf r}}_{pB}$ is the radius-vector connecting $p$ and the center-of-mass of nucleus $B$. 
The first term in Eq. (\ref{reactampl3}) describes the volume integral, in which $r_{pB} \le R$ , while the second term 
determines the contribution of the volume integral in which $r_{pB} \ge R$. We choose the radius $R$ large enough, so that at 
$r_{pB} \ge R$ the UCO ${\hat C}_{pB}=1$. The internal volume integral we can transform into the surface one over the coordinate ${\rm {\bf r}}_{pB}$ using the Green's theorem 
\begin{align}
L = < f({\rm {\bf r}}) |{\overleftarrow T} - 
{\overrightarrow T}| g({\rm {\bf r}})> |_{r \le R}
=  - \frac{1}{2 \mu^{2}}\, R^{2} \nonumber\\
\times \int {\rm d}{\rm {\bf {\hat r}}} 
\left [g({\rm {\bf r}}) \frac{\partial }{{\partial r}} f^{*}({\rm {\bf r}}) -  
 f^{*}({\rm {\bf r}}) \frac{\partial }{{\partial r}} g({\rm {\bf r}})\right ]|_{r=R},
\label{greentheorem1}
\end{align} 
where $\mu$ is the reduced mass of the interacting particles. 
Taking into account Eq. (\ref{greentheorem1}) we can rewrite $M_{dp}$ as  
\begin{align}
M_{dp} ={M_{dp}}^{(S)}|_{r_{pB}= R}  + M_{dp}|_{r_{pB} > R} .
\label{reactampl4}
\end{align}
Here, ${M_{dp}}^{(S)}|_{r_{pB}= R}$ is the reaction amplitude in which the volume integral over ${\rm {\bf r}}_{pB}$ is replaced by the surface integral taken over $\Omega_{{\rm{\bf r}}_{pB}}$ at $r_{pB}=R$. The integrations over all other independent coordinates are carried over without limitations. The operator ${\hat C}_{pB}$ is expressed in terms of the two-body operators $g(p\,j)$, which connect the proton and nucleon $j \in B$.
Each $g(p\,j)$ consists of the radial and tensor parts, which depend on 
${\bf {\rm r}}_{p\,j}=  {\bf {\rm r}}_{pB} - {\bf {\rm r}}_{j}$. Here, ${\rm {\bf r}}_{j}$  i
s the radius-vector connecting nucleon $j$ and the center-of-mass of nucleus $B$.  Since in the final state nucleus $B$ is in a bound state, the integration over all ${\rm {\bf r}}_{j}$,
$\,j \in B$, is limited. Hence at large enough $r_{pB}$ radius $r_{p\,j}$ will also be large enough to exceed the correlation radius, i.e. $g(p\,j)=1$. Thus at large enough $R$ the UCO ${\hat C}_{pB}$ can be replaced by unity, i.e. ${M_{dp}}^{(S)}|_{r_{pB}= R}$ becomes insensitive to the UCOs. The second term, ${M_{dp}}|_{r_{pB} > R}$, is the reaction amplitude in which the volume integral over ${\rm {\bf r}}_{pB}$ is taken from $r_{pB}=R$ and all the integrations over other independent variables are carried out over the whole configurational space. Thus starting from the traditional expression (\ref{ampldp1}) and using the surface integration we have proved that the exact $(d,p)$ reaction amplitude is invariant under the finite-range unitary correlations. 

The invariance of the reaction amplitude is true only if the exact formulation is used, i.e. the exact wave function in the initial state (the post-form) or the exact final state wave function (in the prior-form). The asymptotic behavior of the exact wave function is given by the incident wave plus the elastic, inelastic, rearrangement and breakup scattered waves \cite{kadyrov03}. The amplitude of each scattered wave is the reaction amplitude for transition initial channel $i\, \to $ the final channel corresponding to this scattered wave. For the rearrangement channel $A(d,p)B$  the amplitude of the outgoing wave in the exit channel is the reaction amplitude given by Eq. (\ref{ampldp1}).  The knowledge of the exact wave function for the collision $d+ A$ assumes that we know its asymptotic behavior in all the asymptotic regions \cite{kadyrov03}, i.e. we know the amplitudes of the all opened rearrangement and breakup channels.  

The invariance of the reaction amplitudes under
 the UCOs is understandable: these amplitudes determine the asymptotic behavior of the scattering wave function which is not affected by the finite range UCOs. Although we have 
proved the invariance of the asymptotic behavior of the wave function in the two-fragment channels below we prove that it is also the case for the breakup reactions. 
Thus, the exact reaction amplitude cannot be used to determine unambiguously the SF which is not invariant under UCO. Moreover, the exact reaction amplitude is not parametrized in terms of the SF, which can be calculated as the square of the norm of the overlap function or using the method discussed in \cite{timofeyuk}. 
In contrast to the exact approach, the conventional DWBA is not invariant under UCO. The DWBA is designed to determine the phenomenological SFs by comparing the calculated differential cross section with the experimental one. To express the DWBA amplitude in terms of the SF drastic approximations are done for the initial exact scattering wave function and the transition operator $V_{pB}$. In particular, in the initial scattering wave function only the incident wave is left, while all other opened channels are neglected and coupling to them is taken into account only effectively in terms of the initial distorted wave. The approximation of the initial scattering wave function by the channel wave function $\varphi_{d}\,\varphi_{A}\,\chi_{i}^{(+)}$  in Eq. (\ref{amplcorr2}) leads to zero for the reaction amplitude, in which transition operators are written in terms of the kinetic energy operators: once we write 
the transition operators in terms of the kinetic energy operators, the volume integral can be transformed in to the surface one  and the radius of the surface  should be large enough 
to encircle the volume in which reaction occurs. To calculate the surface integral the correct asymptotic behavior of the exact scattering wave function is required. Moreover as the radius of the surface goes to infinity (for the $(d,p)$ reaction it would be $r_{pB}=R \to \infty$)  only the contribution from the outgoing wave in the exit channel $p+B$ will survive for the $(d,p)$ reaction but this wave is missing in the DWBA.  

Let us write now the DWBA reaction amplitude (in the post-form) taking into account UCOs:
\begin{eqnarray}
M_{dp}^{DW} = <{\tilde \Phi}_{f}^{(-)}\,{\hat C}_{B}^{-1}|\,V_{pB} - U_{pB})|{\hat C}_{pn}\,{\hat C}_{A}\,{\tilde \Phi}_{i}^{(+)}>, 
\label{dwampl1}
\end{eqnarray} 
where we took into account that the channel wave function in the initial state $\Phi_{i}^{(+)}=  \varphi_{d}\,\varphi_{A}\,\chi_{i}^{(+)}= {\hat C}_{pn}\,{\hat C}_{A}\,{\tilde \Phi}_{i}^{(+)}$, $\,{\tilde \Phi}_{i}^{(+)}=  {\tilde \varphi}_{d}\,{\tilde \varphi}_{A}\,\chi_{i}^{(+)}$.
In the adiabatic DWBA (ADWBA) $V_{pB} - U_{pB}$ is replaced by $V_{pn}$.  Then we get the ADWBA amplitude
taking into account UCOs: 
\begin{eqnarray}
M_{dp}^{DW} =  <{\tilde \Phi}_{f}^{(-)}|{\hat C}_{nA}^{-1}\,V_{pn}\,{\hat C}_{pn}|{\tilde \Phi}_{i}^{(+)}>  \nonumber\\
 =<\chi_f^{( - )}\,I_{A}^{B}|V_{pn}\,{\hat C}_{pn}|
{\tilde \varphi}_{d}\,\chi_{i}^{(+)}>.  
\label{dwampl2}
\end{eqnarray}
Since the $(d,p)$ reactions are dominantly peripheral (about $70-80\% $ contribution to the reaction amplitude comes from the nuclear exterior), the probability to find the transferred neutron close to nucleons in $A$ at distances of the range of the UCO is negligible \cite{pang}, i.e. the effect of the correlator ${\hat C}_{nA}^{-1}$ on the DWBA amplitude can be neglected. 
The main impact of the UCO is replacement of $V_{pn}$ by $V_{pn}\,{\hat C}_{pn}$. However, this effect is too small to significantly affect the DWBA amplitude.
 
{\em{Breakup $(d,p\,n)$ reaction amplitude.}}
Here we demonstrate that the deuteron breakup reaction amplitude is also invariant under UCO transformation. Usually to consider the breakup amplitude one starts from the prior-form:
\begin{eqnarray}
M_{dpn} = &  <\Psi_{f}^{(-)}|V_{dA}  - U_{dA}|\varphi_{d} \varphi_{A} \chi_{i}^{( + )}> 
=  <\Psi_{f}^{(-)}\,|{\overrightarrow H}   \nonumber\\
& - E |\, \Phi_{i}^{(+)}> 
=   < \Psi_{f}^{(-)}|-{\overleftarrow T} + {\overrightarrow T}|
\, \Phi_{i}^{(+)}>.  \qquad   \label{breakupamps1}
\end{eqnarray}
Here, evidently, $\Psi_{f}^{(-)}$ is the exact scattering wave function in the final state, which has the three-body incident wave $p+ n+ A$ and  $H\,\Psi_{f}^{(-)}\,=\,E\,\Psi_{f}^{(-)}$, $\,H$ is the Hamiltonian of the system $p+ n+ A$. The total kinetic energy operator can be written as $T= T_{p\,n} + T_{dA}+ T_{A}$, where
 $T_{p\,n}$ and $T_{A}$ are hermitian because in the initial state deuteron and $A$ are bound. Hence, 
\begin{align}
M_{dpn} = & < \Psi_{f}^{(-)}|-{\overleftarrow T}_{dA} + {\overrightarrow T}_{dA}|
 \varphi_{d} \varphi_{A} \chi_{i}^{( + )}>   \nonumber \\
= & <{\tilde \Psi}_{f}^{(-)}{\hat C}^{-1}|-{\overleftarrow T}_{dA} + {\overrightarrow T}_{dA}|{\hat C}_{d} {\hat C}_{A} {\tilde \varphi}_{d} {\tilde \varphi}_{A}  \chi_{i}^{( + )}>   \nonumber \\
= & <{\tilde \Psi}_{f}^{(-)}{\hat C}_{dA}^{-1} |-{\overleftarrow T}_{dA} + {\overrightarrow T}_{dA}| {\tilde \varphi}_{d} {\tilde \varphi}_{A} \chi_{i}^{( + )}>. 
\label{brampl1}
\end{align}
where ${\hat C}={\hat C}_{A}\,{\hat C}_{d}\,{\hat C}_{dA}$, $\,{\hat C}_{d}={\hat C}_{pn}$.
The integral over $T_{dA}$ is not hermitian because the relative $d-A$ motion in the initial and final state is unbound.  Note that we can tag the proton and neutron belonging to the deuteron in the initial state because the wave function in the initial state is antisymmetrized only over exchange of nucleons belonging to nucleus $A$.

We can repeat now the steps used when considering the $(d,p)$ amplitude. We divide the volume integral over ${\bf {\rm r}}_{dA}$ into the internal and external parts, and transform the internal volume integral into the surface one. If the radius of this surface is large 
enough  the UCO ${\hat C}_{dA}= {\hat C}_{pA}\,{\hat C}_{nA}$ can be replaced by the unit operator. Evidently the external volume integral in such a case also doesn't depend on ${\hat C}_{dA}$. Thus we proved that the deuteron breakup amplitude is also invariant under the UCO transformation.  Note that we can include the three-body correlator ${\hat C}_{pnA}$ into ${\hat C}_{dA}$  and it still turns into unity when $r_{dA}$ becomes large enough.  
Since we have proved before the invariance of the reaction amplitudes for the rearrangement channels we can conclude that asymptotic behavior of the exact $\Psi_{i}^{(+)}$ in general is invariant under the UCO transformation. We note that for the nucleon knockout in heavy-ion collisions the invariance of the reaction amplitude under the UCOs can be proved in a similar way.

The DWBA amplitude can be obtained from Eq. (\ref{breakupamps1}) by replacing the exact wave function $\Psi_{f}^{(-)}$ by the three-body channel wave function, which is , in contrast to the two-body case, is not determined 
uniquely and is different in the different asymptotic regions. If we take the channel wave function in the final state as $\Phi_{f}^{(-)}= \chi_{pn}^{(-)}\,\chi_{dA}^{(-)}\,\varphi_{A}= {\hat C}_{pn}\,{\hat C}_{A}\,{\tilde \chi}_{pn}^{(-)}\,\chi_{dA}^{(-)}\,{\tilde \varphi}_{A}$ (or its CDCC extension) and take into account that $V_{dA}$ in the DWBA is replaced by the interaction potential, which depends on the distance between the center-of-mass of $d$ and $A$, then we see immediately that the DWBA amplitude doesn't depend on the UCOs. Thus the DWBA amplitude for the $(d,pn)$ reaction is invariant under UCO transformation. 

{\em{$(e,e'\,p)$  reaction amplitude.}}---
The $B(e,e'\,p)A$ process amplitude in the prior-form is given by (similar to the $(d,p\,n)$ reactions)
\begin{align}
M_{ee'p} = &  <\Psi_{f}^{(-)}|V_{eB}  - U_{eB}|\,\varphi_{B}\,\chi_{i}^{( + )}>  
\nonumber  \\
= & < \Psi_{f}^{(-)}|-{\overleftarrow T} + {\overrightarrow T}|\,\varphi_{B}\,\chi_{i}^{( + )}> .
\label{elbrampl2}
\end{align}
Note that this expression is valid even if the three-body potentials $V_{e\,p,j}$ are included in addition to the two-body potential $V_{e\,p}$.  
Introducing the correlators we get 
\begin{align}
M_{ee'p} = & 
< {\tilde \Psi}_{f}^{(-)}\,{\hat C}_{p\,A}^{-1} {\hat C}_{A}^{-1}
|-{\overleftarrow T} + {\overrightarrow T}|{\hat C}_{p\,A}\,{\hat C}_{A}\,{\tilde \varphi}_{B}  
\chi_{i}^{( + )}>  \nonumber \\
= & < {\tilde \Psi}_{f}^{(-)}
|-{\overleftarrow T}_{e\,B} + {\overrightarrow T}_{e\,B}|{\tilde \varphi}_{B} \chi_{i}^{( + )}>, 
\label{elbrampl11}
\end{align}
where $T= T_{e\,B} +  T_{p\,A} + T_{A}$ and $\chi_{i}^{(+)}$ is the electron distorted wave in the initial channel.  To obtain the last equation we took into account that operators $T_{pA}$ and $T_{A}$ are hermitian because $B$ is in the bound state in the initial channel. Thus the exact $(e,e'p)$ amplitude is invariant relative UCOs. The DWBA will be considered elswhere.

In conclusion, we have demonstrated that the  $(d,p),\,(d,pn)$ and $(e,e'p)$ reaction amplitudes determining the amplitudes of the asymptotic terms of the exact scattering wave functions in the corresponding channels are invariant under finite range UCOs while the SFs are not.  
Moreover, the exact reaction amplitudes cannot be parametrized in terms of the SFs, which should be calculated from the overlap functions rather than from the reaction amplitudes. Hence, the exact reaction amplitudes cannot provide a tool to determine the SFs. Only drastic simplifications of the reaction amplitudes lead to the DWBA amplitudes, which are parametrized in terms of the "SFs" determined by comparison with the experimental data.  However, in reality, these "SFs" are nothing else but proportionality coefficients between the DWBA cross sections and experimental data. The DWBA amplitudes for the $(d,p)$ reactions are sensitive to the finite range UCOs but, due to the peripheral character of the $(d,p)$ reactions, this sensitivity is suppressed, while the DWBA amplitudes for the $(d,pn)$ reactions are invariant under the UCOs.

The work was supported by the U.S. Department of Energy under Grant No. DE-FG02-93ER40773 and DE-FG52-06NA26207, NSF under Grant No. PHY-0852653 and the Australian Research
Council. A.S.K. also acknowledges the support from NSF and the kind hospitality of the Cyclotron Institute
during his visit to Texas A \& M University.


\begin{thebibliography}{9}

\bibitem{lee06} Jenny Lee {\it et al.}, Phys. Rev. {\bf C 73}, 044608 (2006).
\bibitem{kramer} G.J. Kramer, H.P. Blokb, L. Lapikás, Nucl. Phys. {\bf A679}, 267 (2001). 
\bibitem{gade} A. Gade {\it et al.}, Phys. Rev. {\bf  C 77}, 044306 (2008).
\bibitem{feldmeier} H. Feldmeier {\it et al.}, Nucl. Phys. {\bf A632}, 61 (1998).
\bibitem{kadyrov09} A. S. Kadyrov {\it et al.}, Ann. Phys. {\bf  324}, 1516 (2009).
\bibitem{kadyrov03}  A.S. Kadyrov {\it et al.}, Phys. Rev. {\bf A 68}, 022703   (2003).
\bibitem{timofeyuk}     N. K. Timofeyuk,   Phys. Rev. Lett. {\bf 103}, 242501 (2009).
\bibitem{navratil} P. Navratil, J. P. Vary and  B. R. Barrett,  Phys. Rev. {\bf C 62},  054311
(2000);  P. Navratil, W. E. Ormand, Phys. Rev. {\bf C 68}, 034305  (2003); S. Quaglioni,  P. Navratil, Phys. Rev. Lett. {\bf 101}, 092501  ( 2008 ).    
\bibitem{pang}  D. Y. Pang, F. M. Nunes,  A. M. Mukhamedzhanov, Phys. Rev. {\bf C 75}, 024601 (2007). 

\end{thebibliography}
\end{document}